\documentclass[12pt]{article}
\usepackage{graphicx,amssymb,amsmath}
\usepackage[colorlinks, linkcolor=red, anchorcolor=blue, citecolor=blue]{hyperref}
\newcommand{\be}{\begin{equation}}
\newcommand{\ee}{\end{equation}}
\newcommand{\bea}{\begin{eqnarray}}
\newcommand{\eea}{\end{eqnarray}}
\newcommand{\beas}{\begin{eqnarray*}}
\newcommand{\eeas}{\end{eqnarray*}}

\begin{document}
\begin{titlepage}

\vspace*{-24mm}
\rightline{LMU-ASC 50/14}
\vspace{4mm}

\begin{center}

{\Large (A)dS Holography with a Cut-off}

\vspace{4mm}

\renewcommand\thefootnote{\mbox{$\fnsymbol{footnote}$}}
Debajyoti Sarkar\footnote{debajyoti.sarkar@physik.uni-muenchen.de}${}^{1,2}$
\vspace{4mm}

${}^1$
{\small \sl Department of Physics and Astronomy} \\
{\small \sl Lehman College of the CUNY, Bronx NY 10468, USA}

${}^2${\small \sl Arnold Sommerfeld Center, Ludwig-Maximilians-University} \\
{\small \sl Theresienstr. 37, 80333 M\"{u}nchen, Germany}


\end{center}

\vspace{4mm}

\noindent
We find out the smearing/ transfer functions that relate a local bulk operator with its boundary values at a cut-off surface located at $z=z_0$ of the AdS Poincar\'{e} patch. We compare these results with de Sitter counterparts and comment on their connections with corresponding construction for dS/ CFT. As the boundary values can help define the required field theory at $z=z_0$ and encode bulk locality in terms of it, our work can provide key information about holographic RG in the context of AdS/ CFT.
\end{titlepage}
\setcounter{footnote}{0}
\renewcommand\thefootnote{\mbox{\arabic{footnote}}}

\section{Introduction}\label{sec:intro}

Since the discovery of AdS/ CFT \cite{Maldacena:1997re}-\cite{Witten:1998qj} almost 17 years ago, one practical approach has been to apply such correspondence to describe and understand the age old difficulties of quantum theories of gravity in terms of the boundary field theory. Many aspects of these gravity theories have already been understood, and many are still underway. In this paper we focus on the issue of locality and causality of quantum gravity theories or at least for their large $N$ cousins. Because only gauge invariant operators are truly local, the general lore is that for gravity theories there is a contradiction in that the diffeomorphism invariant operators are naturally non-local there. However, for large $N$ or zero Planck length limit the graviton fluctuations are small and one can expect to have a perturbative notion of local bulk operators in terms of local boundary operators.\footnote{In what follows, by locality we will mean the statement of `microcausality' \cite{Dubovsky:2007ac}. Said another way, it just means that two spacelike separated local operators commute between themselves.} Our work is based on numerous previous papers which arose after the proper understanding of Lorentzian version of AdS$_{d+1}$/ CFT$_d$ and the role of the boundary conditions \cite{Balasubramanian:1998sn}-\cite{Bena:1999jv}. Lorentzian version of AdS/ CFT plays a key role here which states that the boundary values of the normalizable part of the bulk field $\phi$ corresponds to a particular boundary operator $\mathcal{O}$. The question then is whether a simple relation between $\phi$ and $\mathcal{O}$ is possible, where the bulk field could be described as (to be understood inside an expectation value calculation)
\[
\phi(x,z)\leftrightarrow\int_{boundary}K(x',z|x)\mathcal{O}(x')
\]
for some kernel $K$. The answer turns out to be yes. Such earlier construction of local bulk operators were successively improved later in a series of papers by Hamilton et al. \cite{Hamilton:2005ju}-\cite{Hamilton:2006fh}. In this approach, the kernel that relates the boundary operators to the bulk is AdS covariant and its structure could be fixed completely from symmetry. Also once one analytically continues the boundary spatial coordinates to complex values, the support of the kernel or the smearing function is over a finite patch of the AdS boundary. This not only makes the computation of bulk correlators and commutators much easier, but in some cases, like in Rindler coordinates or for BTZ black holes, doing so is absolutely necessary \cite{Leichenauer:2013kaa}. The bulk fields considered initially were free scalar fields which were later generalized in two different ways: incorporating higher spins \cite{Kabat:2012hp},\cite{Sarkar:InProgress},  and including $1/N$ corrections to the construction \cite{Kabat:2011rz}-\cite{Kabat:2013wga} (for a similar approach, but approaching from the bulk side see \cite{Heemskerk:2012np},\cite{Heemskerk:2012mn}). Such construction could be thought of as defining the bulk fields in terms of the well-defined boundary theory thereby essentially reducing the problem of quantum gravity to a well-understood exercise. 

In what follows we will use the Poincar\'{e} patch of the AdS space. The metric we will use is 
\begin{eqnarray*}
ds^2=G_{MN}dX^MdX^N=\frac{R^2}{z^2}(\eta_{\mu\nu}dx^\mu dx^\nu+dz^2)\\
\mu,\nu=0,\dots,d-1\quad \eta_{\mu\nu}=\mbox{diag}(-,+,\dots,+)
\end{eqnarray*}
with $R$ being the AdS radius (which we set to 1 from here onward) and $z$ being the radial coordinate.

For completeness, let's note that although one expects the infinite $N$ (super)gravity theories to be essentially local, the smearing construction (and especially its generalization to order by order in $1/N$) is extremely useful to understand how locality breaks down as we go to finite $N$. At infinite $N$, if we have a bulk field $\phi$ of mass $m$ and if it corresponds to a local operator $\mathcal{O}$ of conformal dimension $\Delta$, then for a free theory, we can generically write the bulk operator as (to be understood in the limit $z'\to 0$)
\[
\phi(z,x)=\int_{boundary}dt'd^{d-1}y'K_{\Delta}(z,x|z',x')\mathcal{O}_{\Delta}(t+t',x+iy')
\]
with $K$ being a kernel which is usually a function or distribution of the AdS covariant distance
\[
\sigma(z,x|z^\prime,x^\prime)=\frac{z^2+{z^\prime}^2+(x-x^\prime)^2}{2zz^\prime}
\]
On the other hand for interacting bulk theories, i.e. order by order in $1/N$, constructing local bulk operators requires addition of a tower of higher dimensional multi-trace operators to the definition for free fields. It goes as 
\be\label{1/ncorr}
\phi(z,x)=\int dx'K_\Delta(z,x|x')\mathcal{O}_\Delta(x')+\sum_{l}a_l\int dx' K_{\Delta_l}(z,x|x')\mathcal{O}_{\Delta_l}(x')
\ee
with particular set of coefficients $a_l$.\footnote{Different choices of $a_l$ correspond to different field redefinitions in the bulk. We thank Tom Banks for raising this question.} This gives a clearer indication as to how the locality could break down at finite $N$ as we run out of independent higher dimensional operators \cite{Hamilton:2007wj},\cite{Kabat:2014kfa}.

However, in the present paper we use this smearing function methods with the aim of studying the RG flow of the field theory as we integrate out parts of the bulk and vice versa. This is the so called holographic RG flow as the bulk and boundary theories have different dimensions. We know that the radial direction $z$ of the bulk can be identified with the field theory energy scale, and somehow the RG flow in the field theory is naturally built into the radial evolution in the bulk. A recent take on this subject were made by \cite{Heemskerk:2010hk},\cite{Faulkner:2010jy} but the precise understanding of the nature of cut-off that one needs to use in the bulk or the emergence of locality etc. have not yet been fully accomplished. It is conceivable that if we can extract information about some part of the bulk (cut-off at some timelike surface $z=z_0>0$) from the field theory side (living or defined at the cut-off surface $z=z_0$), it can directly probe the above issue. Luckily the smearing technique is quite custom made for addressing such questions. 

One other place where the AdS/ CFT dictionary at a cut-off surface might find its potential, is the dS/ CFT correspondence \cite{Strominger:2001pn},\cite{Anninos:2011ui}. In this context, one can ask whether it is possible to study smearing functions in dS starting from their AdS counterpart. Although the status of dS/ CFT is still debatable \cite{Dyson:2002nt},\cite{Goheer:2002vf} for various reasons, one can take the construction of a field in dS (at large $N$) in terms of its boundary values as a physical problem and try to push the construction to see how much one can extract out of it. It is well known that the field theory one gets from the asymptotic behavior of dS physics is non-unitary. Hence one can consider both normalizable and non-normalizable boundary values unlike the AdS case. Put another way, in AdS/ CFT smearing construction, only the normalizable modes are considered and they are the ones that correspond to the boundary operators. But one can analytically continue the AdS metric to dS (see (\ref{dS2AdSancont}) later) and study the behavior of the smearing function in dS. However as the $z$ direction of AdS becomes the time direction for dS, picking out normalizable modes in AdS pertains to considering either positive or negative frequency modes for dS. Hence the analytic continuation fails to work here. In fact, recently \cite{Xiao:2014uea} considered the description of bulk fields in dS in terms of field theory operators by constructing appropriate smearing functions. Indeed it was found that the dS bulk operator construction (for e.g. scalar fields of mass $m^2>\left(\frac{d}{2}\right)^2$) involves smearing two copies of operators of dimensions $\Delta$ and $d-\Delta$ respectively. For these reasons the usual dS/ CFT correspondence\footnote{By `usual' correspondence, we will always mean that the dual field theory lives at the conformal boundary.} can't be obtained as the analytical continuation of usual AdS/ CFT as done in smearing methods (see also \cite{Harlow:2011ke}). However for a cut-off AdS/ CFT, the boundary values of both the normalizable and non-normalizable modes should be smeared to get a local bulk operator in AdS. Then such an analytical continuation from cut-off slice AdS/ CFT to a cut-off slice dS/ CFT becomes possible. 

To this end, let's discuss the plan of the paper. In section \ref{sec:1} we compute the smearing functions corresponding to the boundary values of the local bulk field in general dimensional AdS space. We then specialize to massless scalars in AdS$_2$ and do some quick crosschecks of the resulting correlators in the subsequent subsections. Section \ref{sec:3.1} is then devoted to exploring the connections with de Sitter space and dS$_2$ in particular. The main result of this article consists of the results derived in these two sections. We then explore the connections of the field theory that one obtains in terms of these boundary value operators with the `deformed' and `cut-off' CFT in section \ref{conn}. In subsection \ref{sec:3.2}, we comment on the holographic RG and write down a few related expressions to relate our program with the existing holographic RG literature. Here our treatment is more qualitative emphasizing the level of difficulty of this problem and we hope to pursue them further in future work. We conclude in section \ref{concl.}. Finally the appendices collect some of the necessary calculations left out during the main text.

\section{AdS Holography on Cut-off Slice}\label{sec:1}

In this section we investigate how the smearing function representation of a local bulk operator should change as we go from the usual AdS/ CFT to a cut-off AdS spacetime. We find out the smearing function for a local scalar bulk operator in AdS$_{d+1}$, which has been cut-off by a surface located at $z=z_0$, in terms of its boundary values at the surface. One can always take these boundary values to define a field theory at the cut-off surface, thereby obtaining a holographic relation between cut-off AdS theory and a field theory. 

In general for Lorentzian AdS$_{d+1}$/ CFT$_{d}$ correspondence, the equation of motion (EOM) of a free massive bulk scalar $\Phi$ has two independent solutions \cite{Balasubramanian:1998sn}. Their Fourier components are
\[
\Phi^{\pm}(z,x)=e^{-i\omega t+i\mathbf{k}\cdot\mathbf{x}}z^{d/2}J_{\pm\nu}(\sqrt{\omega^2-k^2}z)=e^{iqx}z^{d/2}J_{\pm\nu}(|q|z)
\] 
Here 
\[
\nu=\Delta-\frac{d}{2}=\sqrt{\frac{d^2}{4}+m^2}, \quad q=(\omega,\mathbf{k})\quad\mbox{and}\quad q^2=(k^2-\omega^2)<0
\]
For $\nu\in\mathbb{Z}$, $J_{-\nu}$ is replaced by $Y_\nu$. For simplicity we can also assume $\nu>0$. Near AdS boundary ($z\to 0$) the field has two distinct behaviors
\[
\Phi(z,x)=\frac{\phi_b(x)}{2\nu}z^\Delta+z^{d-\Delta}j(x)
\]
where $j$ and $\phi_b$ are defined via
\begin{eqnarray}\label{jphidefn}
j(x)=z^{-d+\Delta}\Phi(z,x)|_{z\to 0}\qquad\mbox{and}\nonumber\\\phi_b(x)=z^{-2\nu}z\partial_z(z^{-d+\Delta}\Phi)|_{z\to 0}\leftrightarrow\mathcal{O}(x)
\end{eqnarray}
They are respectively non-normalizable and normalizable fall-offs of the bulk field. $j$ acts as the current or source to the boundary CFT operators whereas the $\phi_b$ part has a direct correspondence with the expectation value $\langle\mathcal{O}\rangle$ of dimension $\Delta$. Now once we introduce a cut-off surface, if we define $\phi_{b,cut}$ and $j_{cut}$ in the similar way as in (\ref{jphidefn}) (but at $z_0$), i.e.
\begin{eqnarray}\label{jphidefnco}
j_{cut}(x,z_0)=z^{-d+\Delta}\Phi(z,x)|_{z\to z_0}\qquad\mbox{and}\nonumber\\\phi_{b,cut}(x,z_0)=z^{-2\nu}z\partial_z(z^{-d+\Delta}\Phi)|_{z\to z_0}
\end{eqnarray}
then we get
\begin{equation}\label{reln}
\phi_{b,cut}|_{z_0\to 0}=\phi_b\qquad\mbox{and}\qquad j_{cut}|_{z_0\to 0}=j+\frac{\phi_b}{2\nu}z_0^{2\nu}=j
\end{equation}
Clearly we get the correct $z\to 0$ behavior from here as we take $z_0\to 0$. So at the cut-off slice the bulk field has one component which is the expectation value (vev) of the corresponding boundary operator, but now the `source' has both the usual source term and the vev:
\[
\Phi(z,x)=\frac{\phi_{b,cut}(x,z_0)}{2\nu}z^\Delta+z^{d-\Delta}\left(j_{cut}(x,z_0)-\frac{\phi_{b,cut}}{2\nu}z_0^{2\nu}\right)|_{z_0\to 0}
\]

Now as we expand $\Phi$ in normalizable modes, for usual correspondence, it only involves $\Phi^+$; but in this case, it will involve both $\Phi^{\pm}$ as both modes are normalizable.\footnote{For now we can think of it as solving a boundary value problem. To connect it with the usual AdS/ CFT case, we need to make sure that we put the non-normalizable mode to zero everywhere in the bulk. We will discuss this in appendix \ref{app:B}. However if our purpose is to connect to dS/ CFT, both modes are equally important. We have referred to it as cut-off slice (A)dS/ CFT throughout the paper.} Hence let's write $\Phi$ as a linear combination of the two modes (for now choosing $\nu\notin \mathbb{Z}$)
\begin{equation}\label{modeex}
\Phi(z,x)=\int \frac{d^dq}{(2\pi)^{d-1}}z^{d/2}e^{iqx}\left(\phi_{1,\omega k}J_\nu(|q|z)+\phi_{2,\omega k}J_{-\nu}(|q|z)\right)
\end{equation}
Our goal now is to invert this above relation to find $\phi_{1,\omega k}$ and $\phi_{2,\omega k}$ by using (\ref{jphidefnco}). We will get two equations:
\begin{eqnarray*}
\left(\phi_{1,\omega k}J_\nu(|q|z)z^\nu+\phi_{2,\omega k}J_{-\nu}(|q|z)z^\nu\right)|_{z=z_0}=\int \frac{d^dx'}{2\pi}e^{-iqx'}j_{cut}(x',z_0)\nonumber\\
\left(\phi_{1,\omega k}J_{\nu-1}(|q|z)qz^{1-\nu}-\phi_{2,\omega k}J_{1-\nu}(|q|z)qz^{1-\nu}\right)|_{z=z_0}=\int \frac{d^dx'}{2\pi}e^{-iqx'}\phi_{b,cut}(x',z_0)
\end{eqnarray*}
Solving these equations for $\phi_{1,\omega k}$ and $\phi_{2,\omega k}$ and plugging them back in (\ref{modeex}), we get
\begin{equation}\label{k1k2_1}
\Phi(z,x)=\int d^dx'K_1(x'|x,z,z_0)\phi_{b,cut}(x',z_0)+\int d^dx'K_2(x'|x,z,z_0)j_{cut}(x',z_0)
\end{equation}
where 
\begin{align*}
K_1&=&\int_{\omega>|k|} \frac{d^dq}{(2\pi)^d}e^{iq(x-x')}\frac{\pi z_0^{\nu} z^{\frac{d}{2}}\csc{\nu\pi}}{2}\left(-J_{\nu}(qz_0)J_{-\nu}(qz)+J_{-\nu}(qz_0)J_{\nu}(qz)\right)\nonumber\\
K_2&=&\int_{\omega>|k|} \frac{d^dq}{(2\pi)^d}e^{iq(x-x')}\frac{\pi qz_0^{1-\nu} z^{\frac{d}{2}}\csc{\nu\pi}}{2}\left(J_{1-\nu}(qz_0)J_{\nu}(qz)+J_{\nu-1}(qz_0)J_{-\nu}(qz)\right)
\end{align*}
The results are slightly different for integer values of $\nu$. There we get
\begin{align*}
K_1&=&\int_{\omega>|k|} \frac{d^dq}{(2\pi)^d}e^{iq(x-x')}z_0^{\nu-1} z^{\frac{d}{2}}\frac{\left(J_{\nu}(qz_0)Y_{\nu}(qz)-Y_{\nu}(qz_0)J_{\nu}(qz)\right)}{q\left(J_{1-\nu}(qz_0)J_\nu(qz_0)-Y_{\nu-1}(qz_0)Y_{\nu}(qz_0)\right)}\nonumber\\
K_2&=&\int_{\omega>|k|} \frac{d^dq}{(2\pi)^d}e^{iq(x-x')}z_0^{-\nu} z^{\frac{d}{2}}\frac{\left(J_{1-\nu}(qz_0)J_{\nu}(qz)-Y_{\nu-1}(qz_0)Y_{\nu}(qz)\right)}{J_{1-\nu}(qz_0)J_\nu(qz_0)-Y_{\nu-1}(qz_0)Y_{\nu}(qz_0)}
\end{align*}
We note that in both these cases $K_1$ goes to zero as $z\to z_0$. This is expected because essentially $j_{cut}$ represents the value of the bulk field at the cut-off surface. Also, if we want, we can also express the cut-off slice operators in terms of the true boundary operators as 
\begin{align}\label{bbreln}
&j_{cut}\sim z_0^{-d+\Delta}\int_{t'^2+y'^2< z_0^2}dt'd^{d-1}y'(\sigma z')^{\Delta-d}\mathcal{O}_{\Delta,CFT}(t+t',x+iy')\qquad\mbox{and}\nonumber\\
&\phi_{b,cut}\sim z_0^{d+1-2\Delta}\partial_{z_0}\int_{t'^2+y'^2< z_0^2}dt'd^{d-1}y'(\sigma z')^{\Delta-d}\mathcal{O}_{\Delta,CFT}(t+t',x+iy')
\end{align}

We should note that the choice of boundary conditions made in (\ref{jphidefnco}) is not at all unique. In fact one can very easily modify the construction for more general boundary conditions. The details have been spelled out in appendix \ref{altco}. However, this one has couple of advantages. Being similar to the usual AdS/ CFT case, in this case, the boundary values also have scaling dimensions $\Delta$ and $d-\Delta$. This is quite reminiscent of the case for dS \cite{Xiao:2014uea} where one needed two boundary operators of above dimensions to write down a local field in the bulk. Also, during RG flow, this helps us in relating with the original CFT at the true boundary. Moreover, it also has some technical advantages. For the choice of the boundary condition in appendix \ref{altco} e.g., the smearing function $K_1$ there identically goes to zero as $z_0$ goes to zero for AdS$_{d+1>2}$ cases. This is because, by definition, there we have $\phi_{b,cut}\sim z_0^{\#}\mathcal{O}$ where $\#$ is a positive integer for e.g. massless fields in AdS$_{d+1\geq 3}$.

\subsection{Massless Scalars in AdS$_2$}\label{sec: AdS2}

After obtaining the general results for the smearing functions, we now consider their implications and do some simpler calculations for massless scalars in AdS$_2$.\footnote{Note that the different choices of the boundary conditions in section \ref{sec:1} and appendix \ref{altco} are effectively the same for this case.} More details and crosschecks of this case have been collected in appendix \ref{subapp1}. Here we briefly mention some overall observations. Indeed for $\Delta=d=1$, the calculations are much simpler and it can easily be checked that as $z\to z_0$, the $K_2$ integration simply gives $\Phi=j_{cut}(T,z_0)$. In fact, if we define $z_0=\frac{z}{m}$, we have
\begin{eqnarray}\label{k1k21}
K_1=\int\frac{d\omega}{2\pi\omega}e^{i\omega(T-T')}\sin\left((m-1)\frac{\omega z}{m}\right)\quad\mbox{and}\nonumber\\
K_2=\int\frac{d\omega}{2\pi}e^{i\omega(T-T')}\cos\left((m-1)\frac{\omega z}{m}\right)
\end{eqnarray}

In general the $\omega$ integral goes from 0 to $\infty$, but for AdS$_2$ there's no constraint like $\omega>k$ and in the smearing function we can add in modes with negative frequency \cite{Hamilton:2005ju}. 
The final result for the bulk operator becomes
\begin{eqnarray}\label{AdS2gen}
\phi(T,z)=\frac{1}{2}\left[j_{cut}(T+(m-1)z_0,z_0)+j_{cut}(T-(m-1)z_0,z_0)\right]\nonumber\\
+\frac{1}{2}\int_{T-(z-z_0)}^{T+(z-z_0)}dT'\phi_{b,cut}(T',z_0)
\end{eqnarray}

It is easy to see that this prescription gives rise to correct smearing function for usual AdS$_2$/ CFT$_1$ case as we take $z_0\to 0$ \cite{Hamilton:2005ju}. In this limit, we see from (\ref{jphidefn}) that $\phi_{b,cut}=\partial_z\Phi|_{z=z_0\to 0}=\mathcal{O}$. On the other hand we don't get any contribution from the $j_{cut}$ part as that is non-normalizable.\footnote{Note that the $K_2$ function of the smearing doesn't give us 0 as $z_0\to 0$. We get zero because we don't have any source turned on at the boundary and hence $j_{cut}$ is itself zero. See appendix \ref{app:B} for more details.} Hence in (\ref{k1k2_1}) we no longer have any $K_2$ to consider and $K_1$ becomes
\begin{eqnarray}\label{derads2res}
\Phi(z,T)=\int dT'K_1(z,T|T')|_{z_0\to 0}\phi_{b,cut}=\int dT'\int\frac{d\omega}{2\pi\omega}e^{i\omega(T-T')}\sin\left(\omega z\right)\mathcal{O}(T')\nonumber\\
=\frac{1}{4}\int dT'\left[\mbox{sgn}(T'-T+z)-\mbox{sgn}(T'-T-z)\right]\mathcal{O}(T')=\frac{1}{2}\int_{T-z}^{T+z}dT'\mathcal{O}(T')
\end{eqnarray}
using the fact that
\[
\mbox{Inverse F.T}\left[\frac{\sin(\omega z)}{\omega}e^{i\omega T}\right]=\frac{1}{4}\left[\mbox{sgn}(T'-T+z)-\mbox{sgn}(T'-T-z)\right]
\] 
\begin{figure}
\begin{center}
\includegraphics[width=0.8\textwidth, height=0.3\textheight]{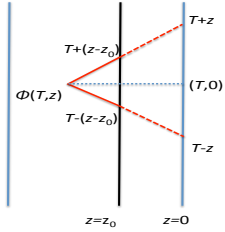}
\end{center}
\caption{Smearing function for massless scalar in AdS$_2$. From (\ref{k1k21}) we see that the support is over the spacelike separated region from the bulk point both at the cut-off slice and at the true boundary.\label{AdS2_case}}
\end{figure}
As mentioned above, we can also check how do the smearing functions behave as we take $z_0\to z$. In this case we should simply recover the bulk field. Indeed for AdS$_2$, the above prescription simply gives $K_1=0$ and $K_2=\delta(T-T')$.

As mentioned before, to compare with the usual AdS/ CFT smearing results, one needs to impose that the bulk field is normalizable everywhere in the bulk and not just for $z_0\to 0$ (as used above). The required condition is given in (\ref{phijreln}) of appendix \ref{subapp1}. Note that for general $z_0$, as shown in appendix \ref{subapp1}, before applying the normalizability condition it is clear that the smearing functions $K_1$ and $K_2$ has support over the spacelike separated (from the bulk point) region on the cut-off surface which goes from $T-(z-z_0)$ to $T+(z-z_0)$ along the boundary time direction (Figure \ref{AdS2_case}). It can be seen easily from (\ref{AdS2gen}).

However, after using the normalizability condition (\ref{phijreln}), the expression can be written as (\ref{expr2}). Also for integer $m$, the usual AdS$_2$ smearing techniques give a particular expression for the bulk field in terms of fields at the cut-off surface. As we have discussed in appendix \ref{subapp1}, (\ref{AdS2gen}) gives rise to that same expected value in that case.

So to summarize, we see that a local bulk operator could be expressed as smearing two sets of operators of scaling dimensions $\Delta$ and $d-\Delta$ (that reside on the cut-off slice, spacelike separated from the bulk point) and we also have the general behavior of the smearing function as we slide the $z_0$ scale.

\subsection{Correlator}\label{corr}

From our construction, the cut-off surface fields that we have smeared in the previous section are nothing but the value of the bulk field at the cut-off surface and its $z$ derivative. We also noted that after imposing the normalizability condition, it properly gives rise to the usual holographic construction of local bulk operators. Therefore it is quite expected that the bulk to bulk correlators and all their local properties will be reproducible starting with the correlators of their boundary values and then smearing them appropriately. We briefly show it here using the simplest example of massless scalar in AdS$_2$ and where $z_0=z/2$. Note that for massive scalars which correspond to a boundary field of dimension $\Delta$, the expression of bulk-to-bulk propagators in AdS$_{d+1}$ is given by (see e.g \cite{D'Hoker:1998mz})
\begin{equation}\label{bbprop}
\langle\phi_\Delta(z,x)\phi_\Delta(z',x')\rangle=2^\Delta\frac{\Gamma(\Delta)\Gamma\left(\Delta-\frac{1}{2}-\frac{d}{2}\right)}{(4\pi)^{\frac{d+1}{2}}\Gamma(2\Delta-d+1)}\sigma^{-\Delta}F\left(\frac{\Delta}{2},\frac{\Delta}{2}+\frac{1}{2},\Delta-\frac{d}{2}+1,\frac{1}{\sigma^2}\right)
\end{equation}
For massless scalars in AdS$_2$, this becomes \cite{Kabat:2011rz}
\begin{equation}\label{bbpropinads2}
\langle\phi_1(z,T)\phi_1(z',T')\rangle=\frac{1}{2\pi}\sigma^{-1}F\left(\frac{1}{2},1,\frac{3}{2},\frac{1}{\sigma^2}\right)=\frac{1}{2\pi}\tanh^{-1} \frac{1}{\sigma}
\end{equation}
These correlators diverge only when the bulk points are lightlike separated or coincident, i.e. when $\sigma=1$. If one of the bulk operator is at the boundary ($z'\to 0$), but null separated from the other, then the regulated distance $\sigma z'\to 0$. This bulk-to-boundary propagator ($\sim \left(\frac{z}{z^2+(x-x')^2}\right)^{\Delta}$) is also divergent at this limit.

As mentioned in appendix \ref{subapp1}, for $\Delta=d=1$ we expect from usual AdS/CFT that for $z_0=z/m$ (for integer $m$),
\begin{align}\label{ads2intm}
\Phi(T,z)&=\sum_{i=1}^{m}\Phi(T_i,z_0)=\sum_{i=1}^{m}j_{cut}(T_i,z_0),\quad \mbox{with}\nonumber\\
&T_1=T+(z-z_0),\quad T_{i+1}=T_i-2z_0,\quad T_m=T-(z-z_0)
\end{align}
Let's study the consequence of this by doing a calculation of two point bulk-bulk correlator and for simplicity, below we choose $m=2$ and $z=z'$. Hence from (\ref{ads2intm}) and (\ref{bbpropinads2}) we expect
\begin{align}\label{2pcorads2}
&\tanh^{-1} \frac{1}{\sigma_0(T',z|T,z)}\nonumber\\
&=\langle (j_{cut}(T'+z/2,z/2)+j_{cut}(T'-z/2,z/2))(j_{cut}(T+z/2,z/2)+j_{cut}(T-z/2,z/2))\rangle\nonumber\\
&=2\tanh^{-1} \frac{1}{\sigma_1(T'+z/2,z/2|T+z/2,z/2)}+\tanh^{-1} \frac{1}{\sigma_2(T'+z/2,z/2|T-z/2,z/2)}\nonumber\\
&+\tanh^{-1} \frac{1}{\sigma_3(T'-z/2,z/2|T+z/2,z/2)}
\end{align}
In fact numerically we indeed find that for two largely separated points, or more precisely when $T'-z'>T+z$ (we are taking $|T'|>|T|$), (\ref{2pcorads2}) is satisfied.\footnote{Note that the AdS$_2$ correlator (\ref{bbpropinads2}) also diverge when $\sigma=-1$ which is the case when $T'-z'=T+z$. It is easily understood that such cases are nothing but a lightcone divergence, as the two operators are related by a light ray coming from the first operator and reflected at the boundary. For $T'-z'<T+z$, (\ref{2pcorads2}) still works as long as it doesn't make any two cutoff surface operators coincident (this will be the case for e.g. $T',z'=1,1$ and $T,z=0,1$).} A more non-trivial check will e.g. be to compute bulk-bulk correlators in AdS$_2$ or higher, starting from the smearing results for non-integer $m$ (\ref{expr2}).

\section{Relation with dS/ CFT}\label{sec:3.1}

After the discovery of AdS/ CFT, it was a natural question as to whether similar correspondences are available for de Sitter spaces too which are important from cosmological perspectives. Indeed there were similar proposals (although string theoretic constructions of dS are harder to come by) by \cite{Strominger:2001pn} and later on higher spin contexts by \cite{Anninos:2011ui}. We will not go into the details of dS/ CFT here, but our focus will rather be an understanding of local operator representation in terms of the dS boundary operators just by assuming a dS/ CFT like correspondence. At least for Poincar\'{e} patch, we will make use of the fact that the AdS metric has a close connection with the flat slicings of dS once one analytically continues the boundary spatial coordinates to complex values. Details of these connections can also be found in the above mentioned papers.

As mentioned earlier, one can't analytically continue the smearing prescription for AdS \cite{Hamilton:2005ju}-\cite{Hamilton:2006fh} to dS space to obtain the correct prescription of dS smearing function. The analytic continuation (done below in (\ref{dS2AdSancont})) flips the $z$ coordinate to dS time coordinates and hence the spacelike separated operators become timelike separated operators. Therefore if we have local bulk operators in AdS, the same prescription can't make it local for dS. Physically, it is because as the $z$ coordinate takes over the role of dS time coordinate (see below), throwing off non-normalizable part pertains to throwing off either the negative or positive frequency modes in dS Cauchy problem. Hence the operators constructed that way will not be local operators in dS. But, the cut-off slice correspondence in AdS/ CFT has a direct connection to dS/ CFT type construction\footnote{In what follows, by dS/ CFT we will refer to at least some techniques that are present in literature, such that, we can smear the corresponding non-unitary field theory operators accordingly to achieve correct correlators in dS space. In general, one considers Hartle-Hawking states in the bulk and the bulk correlators are then required to go over to the Euclidean CFT correlation function. It is in this sense, one should treat a bulk to boundary relation like (\ref{dSresult}) (see also (\ref{dScorr}) later and the discussions following it). However, dS/ CFT has some intuitive and physical problems unlike AdS/ CFT \cite{Dyson:2002nt},\cite{Goheer:2002vf}. There are also alternatives of dS space duality such as dS/ dS \cite{Alishahiha:2004md} or static patch solipsism \cite{Anninos:2011af}, which we won't consider here. It is also interesting to see whether one can construct bulk operators in dS which give correct correlators in the so called $\alpha$-vacua of dS.} as both the components are present. Also as we have shown in appendix \ref{app:1}, the boundary theory corresponding to the (A)dS spacetime decides whether both the operators will survive or not as we take $z_0\to 0$. Recently \cite{Xiao:2014uea} has computed the boundary field theory representation of operators in dS space by assuming the existence of a dS/ CFT correspondence and it indeed turns out that to construct local operators in de Sitter, one needs to smear two sets of boundary operators over the boundary region, timelike separated from the bulk point. It then gives the correct dS two point functions and so forth (corresponding to the Wightman functions in the Euclidean vacuum). The two boundary operators have dimensions $\Delta$ and $d-\Delta$ respectively where the mass of the bulk scalar field is related to the conformal dimension as $\Delta=\frac{d}{2}+i\sqrt{m^2-\left(\frac{d}{2}\right)^2}$. The dS metric (rather its flat slicings) could be related to the AdS metric via the analytical continuation shown below
\begin{eqnarray}\label{dS2AdSancont}
z\to \eta,\quad T\to T, \quad x^i\to ix^i \quad \mbox{and} \quad R_{AdS}\to iR_{dS}\nonumber\\
ds^2_{dS}=\frac{-d\eta^2+d\mathbf{x}^2}{\eta^2}
\end{eqnarray}
Then for bulk operators with $m^2>\left(\frac{d}{2}\right)^2$ one has\footnote{\cite{Xiao:2014uea} sticks to $m^2>\left(\frac{d}{2}\right)^2$ case for scalars and $m^2<\left(\frac{d}{2}\right)^2$ case is relevant only for higher spins in dS. The massless scalar case in dS$_2$ seats in the middle of these two.}
\begin{eqnarray}\label{dSresult}
\Phi(\eta,\mathbf{x})=\frac{\Gamma\left(\Delta-\frac{d}{2}+1\right)}{\pi^{d/2}\Gamma\left(\Delta-d+1\right)}\int_{|x'|<\eta}d^d\mathbf{x'}\left(\frac{\eta^2-\mathbf{x'}^2}{\eta}\right)^{\Delta-d}\mathcal{O}_{\Delta}(\mathbf{x}+\mathbf{x'})\nonumber\\+\frac{\Gamma\left(\frac{d}{2}-\Delta+1\right)}{\pi^{d/2}\Gamma\left(1-\Delta\right)}\int_{|x'|<\eta}d^d\mathbf{x'}\left(\frac{\eta^2-\mathbf{x'}^2}{\eta}\right)^{-\Delta}\mathcal{O}_{d-\Delta}(\mathbf{x}+\mathbf{x'})
\end{eqnarray}
The appearance of two such boundary operators of complementary dimensions is reminiscent of our results in section \ref{sec:1}. Hence we analytically continue our AdS$_2$ coordinates to dS$_2$ to see how does the smearing construction in (\ref{AdS2gen}) compare with the de Sitter result.\footnote{Note that the boundary conditions for fields at the cut-off surface in AdS trivially go over to the dS case.} One of the main reason to stick to dS$_2$ is that for higher dimensional cases, finding a compact support of the smearing functions over a finite region of the boundary is a harder problem. 

So for simplicity let's consider massless scalars (this gives $\Delta=0<d/2$. But that's okay, as in dS, the field theory at the boundary doesn't need to be unitary) and take $T=0$. One can follow the derivation of the smearing functions given in \cite{Xiao:2014uea} (it uses retarded Green's function method to compute the smearing, instead of the mode sum approach employed here), to see that we can directly trust the second term of (\ref{dSresult}) even in this case, which precisely boils down to the $K_1$ integration in (\ref{k1k21}) (or the second term of (\ref{AdS2gen})) as we take $z_0\to 0$. The pre-factor $\frac{\Gamma\left(\frac{d}{2}-\Delta+1\right)}{\pi^{d/2}\Gamma\left(1-\Delta\right)}$ in (\ref{dSresult}) also correctly gives $\frac{1}{2}$. 

However the first term in (\ref{dSresult}) diverges due to the gamma function and should not be trusted. However the calculation for this part is precisely equal to the case for Maxwell fields in AdS$_2$ where $\Delta=d-1=0$ \cite{Kabat:2012hp}. This is natural, as the first part of (\ref{dSresult}) is simply the AdS analog of the scalar field case. From appendix A of \cite{Kabat:2012hp}, we know that the corresponding smearing function is of the form $\frac{1}{2}\left[\delta(T'+\eta)+\delta(T'-\eta)\right]$. But this is precisely the first term of (\ref{AdS2gen}) after analytic continuation. This shows explicitly that the smearing formulas for cut-off slice dS$_2$ are direct analytic continuation of cut-off slice AdS$_2$ case.\footnote{Note that the main difference between taking $z_0\to 0$ limit in AdS and dS is that for AdS, we threw away the $K_2$ part in (\ref{k1k21}) as there it plays the role of a boundary source term which can be taken to zero. In dS, it's essential to path integrate over the sources too.}

\section{Field Theory at the Cut-off Surface}\label{sec:reln}

Because we are constructing bulk operators in terms of operators at the cut-off slice, our approach is quite custom-made to answer questions of holographic RG flows. The understandings of holographic RG were made sharper by \cite{Heemskerk:2010hk},\cite{Faulkner:2010jy} and so on where the flow equations were obtained in the functional form by properly defining the IR and UV part of the bulk wavefunction and then studying their radial evolution equations. The general idea is that integrating out the bulk spacetime induces multitrace deformations to the original CFT and a boundary action at the cut-off surface was also identified which defines the boundary conditions for the bulk fields. Below we relate our construction with such `Wilsonian' flow.\footnote{It should be noted that there are various intuitive subtleties with such constructions. E.g. in Wilsonian RG, one is supposed to keep all the bulk fields, whereas holographic RG compels one to throw away bulk fields for some radial values. Other subtleties involve massless modes in the integrated out region, high energy modes in the IR region and so on.}

\subsection{Connection with Cut-off CFT and Deformed CFT}\label{conn}

Before relating the two constructions, we first clarify two potential confusions. Firstly, our goal here is to stick to a local bulk description even for the cut-off region of the bulk. Such description is easily found by promoting the cut-off surface boundary values $j_{cut}$ and $\phi_{b,cut}$ to operators and then defining a field theory in terms of their correlators. However, it is not immediately clear whether they are also the operators that should exist in the dual cut-off CFT (cut-off CFT in the sense of \cite{Heemskerk:2010hk},\cite{Faulkner:2010jy} etc. where the $z_0$ surface acts as a UV cut-off to the CFT. For an expression see e.g. (\ref{psiir}) later) or whether the cut-off CFT can already give rise to a local bulk description. We can clarify this confusion by looking at bulk lightcone divergences of the local bulk correlators in the smearing constructions, where the bulk lightcone divergences precisely come from the CFT UV divergences which occur at the points where the bulk lightcones hit the boundary. But when we cut off some part of the bulk, if we claim that a local bulk operator is possible to construct by smearing some cut-off conformal field theory operators, we see that we run into a contradiction. Hence the $\phi_{b,cut}=\tilde{\mathcal{O}}_{\Delta}$ and $j_{cut}=\tilde{\mathcal{O}}_{d-\Delta}$ operators (the tilde's are given to distinguish them from true boundary operators) are not really the operators present in the cut-off CFT. The true cut-off CFT operators, if smeared and integrated on a finite slice of the cut-off surface - spacelike separated from the bulk point - give us a description of bulk operators which are not strictly local (the non-locality is presumably over length scale $z_0$). Thus we avoid the bulk lightcone singularities. However the field theory defined by $j_{cut}$ and $\phi_{b,cut}$ is a local field theory and thus can gladly incorporates the locality properties of bulk correlators, namely the lightcone divergences and so on. 

Secondly, the local field theory defined by the correlators of $j_{cut}$ and $\phi_{b,cut}$ (from here on denoted as LFT) can't also be related to any deformation of the boundary CFT by multi-trace (for simplest case, consider double-trace) deformations (from here on denoted as dCFT). It can be seen in a couple of ways: as a first step, we can compute the correlation functions of $j_{cut}$'s  for $z_0$ close to 0 and try to see whether one can deform the CFT action to obtain such correlators. Expanding correlators of $j_{cut}$ and $\phi_{b,cut}$ for near boundary points are apparently an easier route as the correlators of $j_{cut}$'s are nothing but bulk-bulk correlators.\footnote{Note that once we use the normalizability conditions of bulk modes (thus relating $\phi_{b,cut}$ with $j_{cut}$ as in (\ref{phijreln})), we can essentially write everything in terms of $j_{cut}$ correlators.} Hence, we start with the bulk-bulk correlators with both points very close to the boundary, i.e. large $\sigma$. For simplicity in the case of AdS$_2$, upon expanding the hypergeometric function in the correlator (\ref{bbpropinads2}) to first order we get (we should keep in mind that here $z,z'\to 0$ and in the expansion, instead of $\sigma$, one needs to use the regulated distance $\sigma zz'$. This way in the boundary limit, they correspond to correlation function of dimension-1 operators and not dimension- 0 operators)
\begin{equation}\label{correxp}
\langle\phi_1(z,T)\phi_1(z',T')\rangle=\frac{1}{2\pi}\left(\sigma^{-1}+\frac{1}{3\sigma^3}+\frac{1}{5\sigma^5}+\dots\right)
\end{equation}
The leading order term here gives back the usual CFT correlation function, whereas the second term indicates that the CFT needs to be deformed in such a way such that the correlation function is modified by $\frac{1}{6\pi}\frac{1}{(T-T')^6}$. Such correlation function appears due to the presence of two single trace operators of dimension-3 at the boundary and at a particular time. The later terms denote correlation function of dimension 5, 7, 9 operators respectively. This Taylor series expansion of bulk correlators is the same as what we would obtain if we Taylor expand the boundary operators around small $T'$ in e.g.\footnote{As we take $z_0\to 0$, i.e. as the perturbation expansion of hypergeometric function gets better and better, the approximation of the bulk operator as a boundary operator at a point gets better and better. In fact as $z_0\to 0$, to leading order, the bulk operator is just the boundary operator $\mathcal{O}(\tilde{T})$.}
\[
\phi(T,z)=\frac{1}{2}\int_{T-z}^{T+z}dT'\mathcal{O}(T')=\frac{1}{2}\int_{T-z}^{T+z}dT'\left[\mathcal{O}(\tilde{T})+(T'-\tilde{T})\partial_{\tilde{T}}\mathcal{O}(\tilde{T})+\dots\right]
\]
and then compute the correlator. The dimension- 3 single trace boundary operators mentioned above are nothing but the $\frac{1}{2}(T'-\tilde{T})^2\partial^2_{\tilde{T}}\mathcal{O}(\tilde{T})$ appearing above (which are then integrated over some time interval $T'$). The terms which correspond to dimension even operators always come with odd powers of $(T'-\tilde{T})$, which when integrated against the smearing function (symmetric under the time coordinates) gives zero in (\ref{correxp}). 

But now we see the problem if we want to compare these perturbative correlator structure with some dCFT correlator structure. The LFT can't be derivable from a multi-trace deformation, because of the powers of order $\frac{1}{N}$ difference between the single trace and multi trace correlators. 

\subsection{Connection with Holographic RG}\label{sec:3.2}

Moreover, even though we argued in section \ref{sec:3.1} that our construction in AdS now bears direct analytic continuation to dS, the situation could be a bit confusing because of the argued inequivalences of dS/CFT and AdS/CFT prescriptions in terms of holographic RG flow languages \cite{Harlow:2011ke}. Here by inequivalence we mean that even though for AdS the GKPW dictionary \cite{Gubser:1998bc},\cite{Witten:1998qj} and BDHM dictionary \cite{Banks:1998dd}  are equivalent, for dS they are not. It can be most easily realized by the fact that even though the AdS correlation functions (by operator insertions at the cut-off slice e.g) are given by 
\begin{equation}\label{AdScorr}
\langle\tilde{\phi}(x_1,z_0)\dots\tilde{\phi}(x_n,z_0)\rangle_{AdS}=\int_{z=z_0}\mathcal{D}\tilde{\phi}\Psi_{IR}[\tilde{\phi}]\tilde{\phi}(x_1,z_0)\dots\tilde{\phi}(x_n,z_0)\Psi_{UV}[\tilde{\phi},\phi_b],
\end{equation}
for dS the correct expression is 
\begin{equation}\label{dScorr}
\langle\tilde{\phi}(x_1,\eta_0)\dots\tilde{\phi}(x_n,\eta_0)\rangle_{dS}=\int_{\eta=\eta_0}\mathcal{D}\tilde{\phi}\Psi^*_{E}[\tilde{\phi}]\tilde{\phi}(x_1,\eta_0)\dots\tilde{\phi}(x_n,\eta_0)\Psi_{E}[\tilde{\phi}]
\end{equation}
Here we have used notations of \cite{Heemskerk:2010hk},\cite{Harlow:2011ke}, namely the tilde fields correspond to values of the fields at the slice $z=z_0$ and $\Psi_E$ is the analytic continuation of the AdS IR wavefunction $\Psi_{IR}$ to the dS space (which is the Hartle-Hawking vacuua for dS space and also Bunch-Davies vacuua for the flat slicings). Here in computing the equal time correlation functions in dS, we pick a vacuum and time evolve to find the associated wave function at time $\eta=\eta_0$ and the integration is over all the field values at $\eta=\eta_0$. The problem is that $\Psi_{UV}$ and $\Psi^*_{E}$ are different and not related by analytic continuation.

But the confusion straightens out once we realize that we were not doing the same thing in section \ref{sec:3.1}. In terms of correlation functions, what we were doing there was writing a bulk correlator in AdS in terms of smearing the corresponding cut-off surface correlator. This can be directly analytically continued to dS space and there by taking $\eta_0\to 0$, we recover the connection between bulk Wightman function with the past or future boundary (non-unitary) CFT correlators of flat sliced dS.\footnote{So in some sense it is an analytic continuation of BDHM dictionary for cut-off surface AdS to cut-off surface dS. On the other hand, the inequivalence we mentioned earlier can be thought of as the fact that even though the GKPW prescription of AdS can be analytically continued to GKPW of dS, one can't do so for BDHM prescription.} As mentioned previously, away from boundary, it is the dS boundary value problem and AdS boundary value problems that are related and the interpretation fails once the normalizability condition is imposed.

\begin{figure}
\begin{center}
\includegraphics[width=0.6\textwidth, height=0.3\textheight]{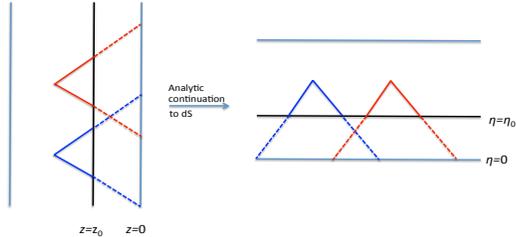}
\end{center}
\caption{Analytic continuation of e.g. two-point bulk correlators from cut-off slice AdS to cut-off slice dS could be possible in terms of holographic RG language. However the connection breaks down for full dualities.\label{AdS_to_dS}}
\end{figure}

Hence the discussion in section \ref{sec:3.1} indicates that an holographic RG type statement which connects cut-off slice AdS/ CFT (before imposing normalizability condition everywhere in the bulk, but only at the boundary) to cut-off slice dS/ CFT by a simple analytic continuation is possible (in a functional integral language), although such equivalence breaks down for the full dualities (figure \ref{AdS_to_dS}). But of course we should note that such a statement for dS is nothing but a statement of bulk time evolution.

Finally, below we write out few expressions to connect our constructions with the current prescription of holographic RG. We begin by briefly recalling the construction of \cite{Heemskerk:2010hk} which, like Wilsonian RG, separate the bulk path integrals into $z>z_0$, $z<z_0$ and $z=z_0$ part.\footnote{In \cite{Heemskerk:2010hk}, $z_0$ has been denoted by $\ell$. Also $\mathcal{O}_i$'s are a complete set of local single trace operators built from the matrix fields and their derivatives and $\kappa\sim\sqrt{G_{Newton}}$.} The bulk path integral can then be written as 
\[
Z=\int \mathcal{D}\tilde{\phi}\Psi_{IR}(z_0,\tilde{\phi})\Psi_{UV}(z_0,\tilde{\phi})
\] 
where $\Psi_{IR}$ and $\Psi_{UV}$ arises from integrating out relevant bulk fields against the exponential of the relevant part of the bulk action:
\[
\Psi_{\stackrel{IR}{UV}}=\int\mathcal{D}\phi|_{\stackrel{z>z_0}{z<z_0}}e^{-\kappa^{-2}\mathcal{S}|_{\stackrel{z>z_0}{z<z_0}}}
\]
As has been postulated in there, if we take\footnote{As mentioned in \cite{Heemskerk:2010hk}, the integration here is over the boundary fields $M$ with a cut-off scale $\delta$, which for pure AdS is simply $z_0$. Also these $j_{cut}$'s have been denoted by $\tilde{\phi}$. Even if we're denoting them to be $j_{cut}$, we need to remember that the integral over $\tilde{\phi}$ appearing in the full bulk path integral below, is over the off-shell fields.}
\begin{equation}\label{psiir}
\Psi_{IR}(z_0,j_{cut})=\int\mathcal{D}M|_{k\delta<1}\exp\left\{-S_0+\frac{1}{\kappa^2}\int d^dxj_{cut}\mathcal{O}_{i}\right\}
\end{equation}
and if we consider the UV factor is a local Gaussian for a single bulk scalar, i.e.
\[
\Psi_{UV}(z_0,j_{cut})=\exp\left\{-\frac{1}{2h\kappa^2}\int d^dx(j_{cut}(x,z_0)+g(x))^2\right\}
\]
Then after the UV part is integrated out we are left with the bulk partition function
\[
Z\propto\int\mathcal{D}M|_{k\delta<1}\exp\left\{-S_0-\frac{1}{\kappa^2}\int d^dx(g(x)\mathcal{O}(x)-\frac{h}{2}\mathcal{O}(x)^2)^2\right\}
\]
This is still the bulk partition function, but it precisely indicates that once one starts integrating out the bulk or equivalently high frequency boundary modes, one induces multi-trace deformation terms in the CFT. This is what we denoted by dCFT before. 

However $\Psi_{IR}$ is to be interpreted as the wavefunction for the cut-off part of the bulk. From that point of view, one can simply treat the $j_{cut}$'s appearing on the exponential of $\Psi_{IR}$ to be the on-shell operators and the $\mathcal{O}_i$'s as their sources\footnote{One can think of it as a `dual' description as in Legendre transformed expressions in quantum field theory. While going from the generating function $W$ to effective action $\Gamma$, $j$ and $\langle\mathcal{O}\rangle$ switch roles. We thank Ivo Sachs for discussions on this point.}. 
Then according to the prescription of \cite{Heemskerk:2010hk}, we have\footnote{Here we write the integration variable as $\tilde{\phi}$, to emphasize that they are off-shell.}
\begin{equation}\label{hrgconn}
\frac{\delta}{\delta\mathcal{O}}\frac{\delta}{\delta\mathcal{O}}Z_{dCFT}|_{\mathcal{O}\to 0}=\frac{\delta}{\delta\mathcal{O}}\frac{\delta}{\delta\mathcal{O}}Z|_{\mathcal{O}\to 0}=\int \mathcal{D}\tilde{\phi}\langle j_{cut}j_{cut}\rangle\Psi_{UV}[\tilde{\phi}]\propto \langle j_{cut}j_{cut}\rangle
\end{equation}
Integrating over $\tilde{\phi}$ for generic Gaussian $\Psi_{UV}$ just gives some factors of $h$ and $\kappa$. Thus one can simply recover the cut-off bulk locality or relate the LFT correlators with the bulk partition function while flowing in holographic RG. 

\section{Conclusions and Outlooks}\label{concl.}

The goal of our paper was mainly twofold. From the point of view of solving a boundary value problem, it is completely conceivable that by explicitly smearing two sets of operators (of some particular scaling dimensions) residing at a cut-off surface, one can always reproduce a corresponding local bulk operator. In turn, it then helps us to reflect upon both the connection of AdS/ CFT with dS/ CFT and also the issues of holographic RG for AdS/ CFT. In general an introduction of such cut-off surface could be quite complicated as the gravity is dynamic at a surface of constant $z$ \cite{Randall:1999vf} and the free field mode expansions should never work when there is such a surface. However, we have totally side-tracked those issues by staying at the large $N$ limit. It would then be quite interesting to extend these calculations by taking $1/N$ corrections into account. Even for usual AdS/ CFT, only the presence of a tower of higher dimensional multi-trace operators, as in (\ref{1/ncorr}), could revive back locality. But now to study bulk locality during RG flow, we probably need another layer of multi-trace operators to deform the CFT with, to eventually relate it to the dual field theory at the cut-off surface. 

There are a few technical barriers that we didn't overcome in this article. We haven't compactified the cut-off surface smearing functions for higher dimensional AdS spaces and also didn't explore the connections fully with higher dimensional dS cases. For compact supports in higher dimensional AdS cases, it is probably required to continue the boundary spatial coordinates again to the complex values. We don't see any reasons to believe that the conclusions we found for two dimensional bulk will be any different for higher dimensions, but still that would be an interesting problem to solve. One can also try to extend this construction for higher spin (HS) fields as they ultimately play crucial roles for 4 and higher point functions. It should be a fun exercise to extend the construction of HS fields for usual (A)dS/ CFT \cite{Kabat:2012hp}, \cite{Sarkar:InProgress}, \cite{Xiao:2014uea} again to this case. However, even for usual dS/ CFT, HS construction is far more subtle than the AdS counterpart and this is a topic currently in progress \cite{Sarkar:InProgress}.\footnote{Remembering that the equations for bulk gauge fields in `holographic gauge' (where all the $z$-components of the bulk fields are taken to zero) just become massless or massive scalar equations (also true for spins higher than 2) in AdS, their construction in terms of the cut-off surface operators is quite straight forward using the result of this paper. This cut-off surface construction in AdS and their analytic continuation to dS can also address the HS spin field construction in dS which behave as scalars of mass $m^2<\left(\frac{d}{2}\right)^2$. This is otherwise difficult to construct \cite{Xiao:2014uea}.}  

At the end, the connection to holographic RG and especially the question ``what cut-off on the field theory corresponds to a radial cut-off in the bulk?'' asked in \cite{Heemskerk:2010hk} still remains unanswered, but this construction clearly indicates an alternate approach: one can always write down a local non-conformal field theory which is sufficient to describe the local bulk operator. The main question then is whether there is any way to connect it to a deformation of the boundary CFT or a cut-off CFT. From our result (\ref{hrgconn}), we see that indeed one can make use of the role of $\Psi_{IR}$ in holographic RG, to always obtain the local field theory necessary for cut-off bulk locality from the bulk partition function. Although we haven't given a clear identification of the nature of UV cut-off in the field theory when putting an IR cut-off in the bulk (and vice-versa), we hope this investigation will lead a way to clarify the above-mentioned confusions, at least when applied to some simplified but particular AdS/ CFT models.

So far, the smearing construction have been able to provide answers to many perturbative and non-perturbative questions regarding the quantum gravity theories. There are still various other questions that one can ask and that is in principle addressable through this technique. One such question is investigating the properties of dual field theory to have a local bulk dual. Many papers already exist which approach the problem from the perspective of conformal bootstrap \cite{Heemskerk:2009pn},\cite{Fitzpatrick:2012yx}, but it can also be addressed from smearing point of view \cite{Sarkar:InProgress2}. Problems such as background independence of the construction or flat space limits are also very deep and interesting. Our present construction of smearing functions on a cut-off surface is a direction where all these questions can again be asked. Lastly, we will like to point out the structural (and visual) similarities of our cut-off slice correspondence techniques and diagrams with the tensor network structures. Recently, tensor networks and the related studies of coarse graining the bulk via multi-scale entanglement renormalization ansatz (MERA) have been a topic of persistent interests \cite{Swingle:2012bk}. They not only address the issues of holographic RG flow, but they originally arose from the topics of condensed matter physics. Hence, it remains to be seen if the current approach towards bulk locality via smearing function can help unify and improve both the topics or not. 

\vspace{1cm}
\centerline{\bf Acknowledgements}
\noindent
We are grateful to Bart Horn, Ivo Sachs and Xiao Xiao for valuable conversations and especially to Dan Kabat for various illuminating discussions throughout the project.  This work was supported in part by U.S.\ National Science Foundation grants PHY-0855582 and PHY11-25915 and by PSC-CUNY and ERC Self-completion grants.


\vspace{1cm}
\centerline{\bf Appendix}
\noindent

\appendix

\section{More General Boundary Condition at the Cut-off Surface}\label{altco}

As advertised in section \ref{sec:1}, the choice of the boundary condition (\ref{jphidefnco}) is not at all unique. In fact, one can easily consider more general boundary values without any pre-factors of powers of $z$ in front. E.g. we can choose (we use the tilde's to distinguish with the choice of section \ref{sec:1})
\begin{equation}\label{bc2}
\Phi(z,x)|_{z=z_0}=\tilde{j}_{cut}(x,z_0)\quad\mbox{and}\quad\partial_z\Phi(z,x)|_{z=z_0}=\tilde{\phi}_{b,cut}(x,z_0)
\end{equation}
These boundary values then correspond to two bulk operators of scaling dimensions 0 and 1 respectively. This is precisely equivalent to knowing the position and velocity in order to solve for a second order differential EOM. Our goal now is to invert (\ref{modeex}) to find $\phi_{1,\omega k}$ and $\phi_{2,\omega k}$ by using the above boundary conditions. Once again, we will get two equations:
\begin{align*}
&\left(\phi_{1,\omega k}J_\nu(|q|z)z^{\frac{d}{2}}+\phi_{2,\omega k}J_{-\nu}(|q|z)z^{\frac{d}{2}}\right)|_{z=z_0}=\int \frac{d^dx'}{2\pi}e^{-iqx'}\tilde{j}_{cut}(x',z_0)\quad\mbox{and}\nonumber\\
&\left[\phi_{1,\omega k}z^{d/2-1}\left(J_{\nu-1}(|q|z)qz+(d-\Delta)J_{\nu}(|q|z)\right)+\phi_{2,\omega k}z^{d/2-1}\left(J_{-1-\nu}(|q|z)qz+\Delta J_{-n}(|q|z)\right)\right]|_{z=z_0}\nonumber\\
&=\int \frac{d^dx'}{2\pi}e^{-iqx'}\tilde{\phi}_{b,cut}(x',z_0)
\end{align*}
Once again, solving these equations for $\phi_{1,\omega k}$ and $\phi_{2,\omega k}$ 
and plugging them back in (\ref{modeex}), we get
\begin{equation}\label{smearcutoff}
\Phi(z,x)=\int d^dx'K_1(x'|x,z,z_0)\tilde{\phi}_{b,cut}(x',z_0)+\int d^dx'K_2(x'|x,z,z_0)\tilde{j}_{cut}(x',z_0)
\end{equation}
where 
\begin{align*}
&K_1=\int_{\omega>|k|} \frac{d^dq}{(2\pi)^d}e^{iq(x-x')}\frac{\pi z_0^{1-d/2} z^{\frac{d}{2}}\csc{\nu\pi}}{2}\left(J_{-\nu}(|q|z_0)J_{\nu}(|q|z)-J_{\nu}(|q|z_0)J_{-\nu}(|q|z)\right)\nonumber\\
&K_2=\int_{\omega>|k|} \frac{d^dq}{(2\pi)^d}e^{iq(x-x')}\frac{\pi z_0^{-d/2} z^{\frac{d}{2}}\csc{\nu\pi}}{2}\nonumber\\
&\Bigg[\left(qz_0J_{-1+\nu}(|q|z_0)+(d-\Delta) J_{\nu}(|q|z_0)\right)J_{-\nu}(|q|z)-\left(qz_0J_{-1-\nu}(|q|z_0)+\Delta J_{-\nu}(|q|z_0)\right)J_{\nu}(|q|z)\Bigg]
\end{align*}
For $\nu\in\mathbb{Z}$ we have
\begin{align*}
&K_1=\int_{\omega>|k|} \frac{d^dq}{(2\pi)^d}e^{iq(x-x')}\left(\frac{z}{z_0}\right)^{\frac{d}{2}}\frac{1}{A}\left(z_0J_{-\nu}(|q|z_0)J_{\nu}(|q|z)-z_0J_{\nu}(|q|z_0)J_{-\nu}(|q|z)\right)\nonumber\\
&K_2=\int_{\omega>|k|} \frac{d^dq}{(2\pi)^d}e^{iq(x-x')}\left(\frac{z}{z_0}\right)^{\frac{d}{2}}\frac{1}{A}\nonumber\\
&\Bigg[\left(qz_0J_{-1+\nu}(|q|z_0)+(d-\Delta) J_{\nu}(|q|z_0)\right)J_{-\nu}(|q|z)-\left(qz_0Y_{-1+\nu}(|q|z_0)+(d-\Delta) Y_{\nu}(|q|z_0)\right)J_{\nu}(|q|z)\Bigg]
\end{align*}
with 
\[
A=qz_0J_{\nu-1}(|q|z_0)J_{-\nu}(|q|z_0)+J_{\nu}(|q|z_0)(-qz_0Y_{\nu-1}(|q|z_0)+(d-\Delta)(J_{-\nu}(|q|z_0)-Y_{\nu}(|q|z_0)))
\]
We note that in this case too, $K_1$ goes to zero as $z\to z_0$. One can also check that the smearing functions obtained here are compatible with the ones in section \ref{sec:1}, by noting that the boundary conditions (\ref{jphidefnco}) and (\ref{bc2}) are same for massless scalars in AdS$_2$ (considered in section \ref{sec: AdS2} e.g). For this case, comparing $K_1$ and $K_2$ with (\ref{k1k21}) e.g, we see that they match up.

\section{Massless Scalar Field in AdS$_2$}\label{app:B}

In section \ref{sec:1} we calculated the smearing functions for the two boundary values of the bulk field in order to obtain back the local bulk operator. As mentioned earlier, this computation is especially important to relate to the results for de Sitter case. But to connect it with the usual AdS/ CFT we need to make sure that the non-normalizable behavior of the bulk field is zero everywhere and especially at the boundary. Below in section \ref{app:1}, we first find out the minimum conditions required to connect our results with usual AdS/ CFT, as we take the cut-off surface to the true boundary. We find that as $z_0\to 0$, the conditions are actually different for AdS and dS cases which is governed by the unitarity property of the boundary theory. Finally among other things, in section \ref{subapp1}, we find out the condition that puts the non-normalizable mode to zero everywhere in the bulk.  

\subsection{Boundary Limit of Section \ref{sec:1} Results}\label{app:1}

For a scalar field $\Phi$ in AdS and for boundary conditions given in (\ref{jphidefnco}), we can write the relation between the two types of boundary conditions as 
\[
\phi_{b,cut}(x)=z_0^{1+d-2\Delta}\partial_{z_0}j_{cut}(x)\Rightarrow z_0^{2\Delta}\frac{1}{z_0^{d+1}}\phi_{b,cut}(x)=\partial_{z_0}j_{cut}(x)
\]
at the cut-off surface. Integrating, we get\footnote{Here we are treating $j_{cut}$ and $\phi_{b,cut}$ as if they are $z_0$ independent which is of course not the case. But it doesn't matter. Basically while integrating we are neglecting a term like $\sim z_0^{2\Delta-d}\partial_{z_0}\phi_{b,cut}$. This is effectively fine as long as our discussion concerns the results at the boundary limit. When we take $z_0\to 0$, (\ref{reln}) gives
$\phi_{b,cut}\to \phi_b=\mathcal{O}_{\Delta,CFT}$, which is $z$ independent. So the derivative term in the last line gives zero at the boundary limit.}
\[
j_{cut}\sim\frac{1}{2\Delta-d}z_0^{2\Delta-d}\phi_{b,cut}
\]
Now for AdS dynamics the unitarity bound implies, $\Delta>\frac{d}{2}$. So as $z_0\to 0$ and because $\phi_{b,cut}$ is the normalizable mode $\mathcal{O}$ as we move to $z_0\to 0$ (\ref{reln}), we have $j_{cut}\to 0$ at the boundary. As a result, in both section \ref{sec:1} and appendix \ref{altco} we see that, as we approach the boundary, only the smearing term including $K_1$ gives us back the usual AdS/ CFT results.\footnote{Actually using old holographic renormalization concepts \cite{Skenderis:2002wp},\cite{Fukuma:2002sb} as $z_0=\epsilon\to 0$, we get $z_0^{2\Delta-d}\partial_{z_0}\phi_{b,cut}=\epsilon^{2\Delta-d}\partial_{\epsilon}\epsilon^\Delta\mathcal{O}^{Ren}_{\Delta,CFT}(x)\sim \epsilon^{2\Delta-d}\epsilon^{\Delta-1}$, where $\mathcal{O}^{Ren}_{\Delta,CFT}$ is the renormalized CFT operator. This is then to be integrated over $z_0$.} This argument however imposes that $j_{cut}\to 0$, only as $z_0\to 0$. It doesn't impose normalizability everywhere in the bulk. For that see appendix \ref{subapp1}.

However, this argument will not work for dS/ CFT as there the field theory is non-unitary and we don't have the $\Delta>\frac{d}{2}$ restriction. There, even as we take $z_0\to 0$, we will still have two copies of boundary operators to smear over as done in \cite{Xiao:2014uea}.

\subsection{Connecting with Usual AdS$_2$/ CFT$_1$}\label{subapp1}

\begin{figure}
\begin{center}
\includegraphics[width=0.5\textwidth, height=0.3\textheight]{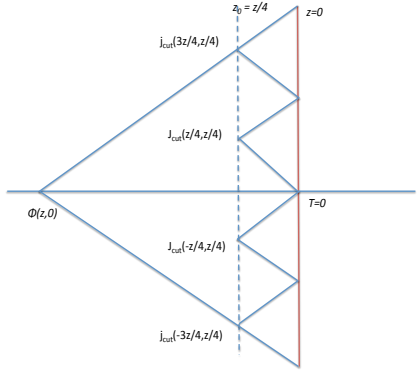}
\end{center}
\caption{Smearing function for AdS$_2$/ CFT$_1$ for cut-off surface $z_0=z/4$ contains four delta functions as shown in the figure. Thus the bulk field $\phi(z,0)$ can be written in terms of four bulk fields all residing at the cut-off surface, but at different times. This is a special case of (\ref{ads2-2}) for $m=4$.\label{AdS2m4}}
\end{figure}

We learned in subsection \ref{sec: AdS2} and appendix \ref{altco} that for a massless scalar in AdS$_2$, i.e. for $\Delta=d=1$, we have
\[
j_{cut}(T,z_0)=\Phi(T,z)|_{z=z_0}\quad\mbox{and}\quad \phi_{b,cut}(T,z_0)=\partial_z\Phi(T,z)|_{z=z_0}=\partial_{z_0}j_{cut}(T,z_0)
\]
and
\begin{eqnarray}\label{ads2}
\Phi(T,z)=\int dT'\left[\int_{-\infty}^{\infty}\frac{d\omega}{2\pi} e^{i\omega(T-T')}\cos[\omega(z-z_0)]\right]j_{cut}(T',z_0)+\nonumber\\
\int dT'\int \frac{d\omega}{2\pi\omega}e^{i\omega(T-T')}\sin[\omega(z-z_0)]\phi_{b,cut}(T',z_0)
\end{eqnarray}
This gave us back the usual smearing prescription for $z_0\to 0$ because in this limit, $j_{cut}\to 0$ and $\phi_{b,cut}\to \mathcal{O}$ (appendix \ref{app:1}). Here we are interested to see whether we get back (see figure \ref{AdS2m4} for a pictorial representation of $m=4$ case)
\begin{eqnarray}\label{ads2-2}
\Phi(T,z)=\sum_{i=1}^{m}\Phi(T_i,z_0)=\sum_{i=1}^{m}j_{cut}(T_i,z_0),\quad\mbox{with}\nonumber\\
 T_1=T+(z-z_0),\quad T_{i+1}=T_i-2z_0,\quad T_m=T-(z-z_0)
\end{eqnarray}
which is a special case for massless scalars in AdS$_2$ with $z_0=z/m$ (for integer $m$).\footnote{The simplest case is when the cut-off surface is halfway to $z$ (m=2) for usual AdS$_2$/ CFT$_1$. There \cite{Hamilton:2005ju} 
\[
\phi(T,z)=\frac{1}{2}\int_{T-z}^{T+z}dT'\mathcal{O}(T')=\frac{1}{2}\left[\int_{T-z}^{0}+\int_{0}^{T+z}dT'\mathcal{O}(T')\right]=\phi(z/2,z/2)+\phi(-z/2,z/2)
\]
As can be easily understood, the one dimensional boundary of AdS$_2$ is responsible for this simplification. For higher dimensional cases such relations are more subtle to come by.}
Back in (\ref{ads2}), we see that the first integral simply gives 
\begin{eqnarray*}
\frac{1}{2}\left[\Phi(T+(m-1)z_0,z_0)+\Phi(T-(m-1)z_0,z_0)\right]\\
=\frac{1}{2}\left[j_{cut}(T+(m-1)z_0,z_0)+j_{cut}(T-(m-1)z_0,z_0)\right]
\end{eqnarray*}
for arbitrary $m$. Therefore e.g. for $m=2$, we have 
\begin{align*}
\Phi(T,z)|_{1^{st} part}&=\frac{1}{2}\left[\Phi(T+z_0,z_0)+\Phi(T-z_0,z_0)\right]\nonumber\\
&=\frac{1}{2}\left[j_{cut}(T+z_0,z_0)+j_{cut}(T-z_0,z_0)\right]
\end{align*}
from the first integral. The second integral of (\ref{ads2}) gives 
\[
\Phi(T,z)|_{2^{nd} part}=\frac{1}{2}\int_{T-(z-z_0)}^{T+(z-z_0)}dT'\partial_{z_0}j_{cut}(T',z_0)
\]
Using Leibniz rule we get
\begin{align*}
\Phi(T,z)|_{2^{nd} part}&=\frac{1}{2}\partial_{z_0}\int_{T-(z-z_0)}^{T+(z-z_0)}dT'j_{cut}(T',z_0)\\
&+\frac{1}{2}j_{cut}(T+(z-z_0),z_0)+\frac{1}{2}j_{cut}(T-(z-z_0),z_0)
\end{align*}
Hence for (\ref{ads2}), we get
\begin{equation}\label{afterleibniz}
\Phi(T,z)=\frac{1}{2}\partial_{z_0}\int_{T-(z-z_0)}^{T+(z-z_0)}dT'j_{cut}(T',z_0)+j_{cut}(T+(z-z_0),z_0)+j_{cut}(T-(z-z_0),z_0)
\end{equation}

However to match up with (\ref{ads2-2}) we need to make sure that we consider only normalizable modes everywhere in the bulk of AdS. In the language of section \ref{sec:1}, it pertains to sending $\phi_{2,\omega k}$ to zero. This relates $j_{cut}$ and $\phi_{b,cut}$ as\footnote{This also shows that at the boundary limit $z_0\to 0$, $j_{cut}\to 0$. Note that the results of appendix \ref{app:1} is a special case of this, namely at the limit $z_0\to 0$.}
\begin{equation*}
\int dT'e^{-i\omega T'}\phi_{b,cut}=\frac{\omega J_{-1/2}(\omega z_0)}{J_{1/2}(\omega z_0)} \int dT'e^{-i\omega T'}j_{cut}=\int dT'e^{-i\omega T'}\omega\cot[\omega z_0]j_{cut}
\end{equation*}
i.e.
\begin{equation}\label{phijreln}
\tilde{\phi}_{b,cut}(\omega,z_0)=\omega\cot \omega z_0\tilde{j}_{cut}(\omega,z_0)
\end{equation}
The second integral of (\ref{ads2}) now becomes\footnote{As should be clear from the context, tilde here denotes the fourier transform and not the tilde fields of appendix \ref{altco}.} 
\begin{align*}
\int \frac{d\omega}{2\pi\omega}\sin [\omega(z-z_0)]e^{i\omega T}\tilde{\phi}_{b,cut}(\omega,z_0)
&=\int \frac{d\omega}{2\pi\omega}\sin [\omega(z-z_0)]e^{i\omega T}\omega\cot \omega z_0\tilde{j}_{cut}(\omega,z_0)\nonumber\\
&=\int dT'\int\frac{d\omega}{2\pi}e^{i\omega(T-T')}\sin [\omega(z-z_0)]\cot \omega z_0j_{cut}
\end{align*}
We can now do the $\omega$ integral of the above, case by case for integer $m$ and verify that the value of the local bulk field matches with (\ref{ads2-2}). The results of $m=2$ and $m=3$ are e.g. given below:
\newline
\newline
$\mathbf{m=2}$: Using 
\[
\mbox{Inverse F.T}[e^{i\omega T}\sin[\omega(z/2)]\cot[\omega z/2]]=\frac{1}{2}\left[\delta(T'-(T-z/2))+\delta(T'-(T+z/2))\right]
\]
we recover from (\ref{ads2})
\[
\phi(z,T)=\left[j_{cut}(T+z_0,z_0)+j_{cut}(T-z_0,z_0)\right]
\]
\newline
\newline
$\mathbf{m=3}$: Using 
\begin{align*}
&\mbox{Inverse F.T}[e^{i\omega T}\sin[\omega(2z/3)]\cot[\omega z/3]]\\
&=\delta(T'-T)+\frac{1}{2}\left[\delta(T'-(T-2z/3))+\delta(T'-(T+2z/3))\right]
\end{align*}
we recover from (\ref{ads2})
\[
\phi(z,T)=\left[j_{cut}(T+(z-z_0),z_0)+j_{cut}(T-(z-z_0),z_0)+j_{cut}(T,z_0)\right]
\]
and so on. This also correctly reproduces the bulk to bulk correlators as shown in section \ref{corr}.


For non-integer $m$, i.e general $z_0$, we can put $\phi_{2,\omega k}=0$ from the outset and find out the expression of the local bulk operator. The final expression becomes 
\begin{eqnarray}\label{expr2}
\phi(z,T)=\int dT'\int \frac{d\omega}{2\pi}e^{i\omega(T-T')}\sin \omega z\sin \omega z_0 j_{cut}\nonumber\\+\int dT'\int \frac{d\omega}{2\pi\omega}e^{i\omega(T-T')}\sin \omega z\cos \omega z_0 \phi_{b,cut}
\end{eqnarray}
From (\ref{expr2}) we see that as we take $z_0\to 0$ we get back the correct boundary prescription of local bulk operators (\ref{derads2res}).

Note that at the stage of (\ref{expr2}), even though the bulk operator is local and we've used $\phi_{2,\omega k}=0$, the spacelike support of the smearing function is not obvious. But this is okay as the argument behind spacelike support comes solely from translating the AdS problem to dS space Cauchy problem. But at the cut-off surface, imposing normalizibility condition spoils this interpretation. Before using the normalizibility condition the spacelike support at the cut-off surface is obvious (as in (\ref{ads2})).\footnote{This is quite similar to the fact discussed in \cite{Xiao:2014uea} in the context of usual dS/ CFT representation where the two sets of boundary operators $\mathcal{O}_\Delta$ and $\mathcal{O}_{d-\Delta}$ are each other's shadow operators. They could be related by a non-local relation and whenever the smearing construction is written in terms of only one set of operator, the support becomes non-compact.}

\providecommand{\href}[2]{#2}\begingroup\raggedright

\endgroup

\end{document}